\documentclass[aps,prl,showpacs,amssymb,amsmath,twocolumn]{revtex4}
\usepackage{graphicx}
\usepackage{bm}

\begin{document}
\title{Gravitational spectral shift caused by Casimir stresses}
\author{V. A. \surname{De Lorenci}}
 \email{delorenci@unifei.edu.br}
\author{L. G. \surname{Gomes}}
 \email{lggomes@unifei.edu.br}
\author{E. S. \surname{Moreira Jr.}}
 \email{moreira@unifei.edu.br}
\affiliation{Instituto de Ci\^encias Exatas,
Universidade Federal de Itajub\'a, 
Av.\ BPS 1303 Pinheirinho, 37500-903 Itajub\'a, MG, Brazil} 

\date{\today}

\begin{abstract}
The linearized Einstein field equations with the renormalized stress tensor of
a massless quantum scalar field as source are solved in the 4-dimensional 
spacetime near an infinite plane boundary. The motion of 
particles and light is investigated disclosing a variety of effects. 
For instance, it is shown that light rays 
initially parallel to the boundary do not bend. On the other hand, under 
the same initial conditions, particles 
will be affected by the presence of the quantum field, being accelerated toward
the plane boundary or departing from it depending on the value of the curvature 
coupling parameter $\xi$. It is suggested that $\xi$ can be set by measuring 
the spectral deviation of a light ray travelling perpendicularly to the boundary.
\end{abstract}


\maketitle


%
%

\paragraph{Introduction. ---}
The Casimir effect is a fascinating feature of quantum field theory in 
non trivial spacetimes. It was long ago predicted \cite{casimir1948}
as an attractive force between two parallel perfectly conducting infinite plates.
Roughly, the effect corresponds to the appearance of stress between 
physical boundaries due to the confinement of quantum fields in 
a limited region of space \cite{milton2001}. 
Not long ago the Casimir effect has experimentally been confirmed 
between conductors \cite{lamoreaux1997,mohideen1998} 
with a statistical precision of about $1\%$.  
Quantum fluctuations of the electromagnetic field
give rise to a constant negative energy density in the space between the plates.
%
Similar behavior also occurs with scalar fields. In
this case the energy density is not uniform for an arbitrary
curvature coupling parameter
$\xi$ \cite{romeo2002,kennedy1980,tadaki1986}. When a single plate
is considered, if the conformal coupling is assumed 
($\xi= 1/6$), symmetry arguments can be used to show that the 
stress tensor is identically nil \cite{dewitt1979}.

Taking energy-momentum densities in Casimir-like systems
\cite{milton2001,bordag2001} 
as source of the gravitational field usually leads to tiny effects \cite{ford2010}, 
which are believed to be non-measurable in laboratory particularly when classical
sources are present.
However such quantum densities must be included as source of gravity, otherwise 
the energy-momentum conservation law would be violated \cite{dewitt1979}.
Moreover there might be circumstances in which these effects could be non-negligible
despite their smallness.
In this letter the general motion of particles and light 
near an infinite plane boundary is 
investigated in the framework of the semiclassical theory for gravity. 
It is shown that light rays initially propagating parallel to the plane boundary
will be unperturbed by the presence of the energy-momentum densities associated with a
massless quantum scalar field. On the other hand, a particle 
will be subjected to an acceleration of quantum origin, which depends 
on the value of the curvature coupling parameter. 
It is argued that $\xi$ can be set by measuring the 
spectral deviation of a light ray travelling near the boundary.
Although the aim of this paper is mainly to address a matter of 
principle measurability of the effects is also considered.

As is well-known semiclassical gravity is 
expected to break down at the Planck scale \cite{davies1982}. However there is
another regime in which this approach may be suspicious, namely, when
fluctuations of the energy-momentum tensor are large compared with its vacuum 
expectation value. This aspect was addressed long ago \cite{ford1982} and since
then discussed by many authors \cite{kuo1993}. 
Different criteria for the validity of the semiclassical theory have also been proposed 
\cite{anderson2003} and the feeling is that the issue requires
further investigation. It is conceivable that the discussion of experimental 
tests such as the one addressed here could shed some light on this matter.

\paragraph{Back-reaction. ---}
We begin with a single plane boundary at $z=0$ in the 4-dimensional 
Minkowski spacetime, 
\begin{equation}
ds^2 = c^2dt^2 - dx^2 - dy^2 - dz^2,
\label{2}
\end{equation}
with $t$, $x$ and $y$ varying from $-\infty$ to $\infty$ whereas $z\ge0$.
Quantization of a massless scalar field $\varphi(t,\vec{x})$ satisfying the 
Dirichlet boundary condition $\varphi(t,x,y,z=0) = 0$ on the boundary
results in the following stress tensor \cite{romeo2002,dewitt1979}
\begin{equation}
\left<T_{\mu\nu}\right>\; = \frac{(6\xi-1)\hbar c}{16\pi^2 z^4} {\rm diag}(1,-1,-1,0).
\label{5}
\end{equation}
This result can also be obtained from the corresponding case where two parallel plates 
are considered and one of them is taken to infinity.
We notice that if the Neumann boundary condition were considered Eq. (\ref{5}) would be
changed by a factor $-1$ \cite{romeo2002}.

Once energy density and pressure warp spacetime, back-reaction 
in general relativity caused by a nonvanishing $\left<T_{\mu\nu}\right>$ 
should be taken into account. Although gravitational effects induced by 
these quantum stresses are expected
to be very small, as they are of order $\hbar$, their influence on the motion of 
test particles (including light) may not be negligible. 
Observing Eq. (\ref{5}) we see that there is a situation in which no gravitational effect appears, 
namely, the case of conformal coupling $\xi=1/6$. 
In what follows we shall study 
back-reaction effects on the motion of particles and light in the spacetime
warped by Eq. (\ref{5}) with arbitrary $\xi$. 

Back-reaction is implemented by feeding 
the Einstein equations with the vacuum expectation value of the stress 
tensor,
\begin{equation}
G_{\mu\nu} = -\frac{8\pi G}{c^4} \left<T_{\mu\nu}\right>.
\label{6}
\end{equation}
In this semiclassical approximation matter fields are
quantum fields while the spacetime metric is classical.
In the case when a material  boundary is considered its mass density will be
neglected. (Otherwise a classical
contribution  should be included in the
total energy-momentum tensor. 
Further comments on this matter are made at the end.)  
 
Following the standard linearization procedure \cite{landau1951,foster2006}
we set $g_{\mu\nu} = \eta_{\mu\nu} + h_{\mu\nu}$, 
where $\eta_{\mu\nu}$ is the Minkowski metric in Cartesian coordinates 
and $h_{\mu\nu} << \eta_{\mu\nu}$ corresponds
to a perturbation of the flat background. 
As Eq. (\ref{5}) shows, $\left<T_{\mu\nu}\right>$ is diagonal, 
time-independent and invariant under Poincar\'e 
transformations in the ($x,y$)-plane.
These aspects imply that the most general line element that solves the back-reaction
problem is 
\begin{equation}
ds^2 = [1 + f(z)](c^2 dt^2 - dx^2 - dy^2) - dz^2.
\label{4}
\end{equation} 
Then Eqs. (\ref{6}) lead to
\begin{equation}
ds^2 = 
\left(1+\frac{\alpha}{z^2}\right)(c^2 dt^2 - dx^2 - dy^2) - dz^2,
\label{35}
\end{equation} 
where
\begin{equation}
\alpha \doteq \frac{1-6\xi}{12\pi}l_P^2,
\label{20}
\end{equation}
and $l_P=(\hbar G/c^3)^{1/2}$.
In the above result nonphysical solutions were discarded by a convenient choice of the 
integration constants. It should be noted that for the conformal coupling Eq. (\ref{35}) 
reduces to the Minkowski line element, as it should.

%
%
\paragraph{ Motion of particles and light. ---}
General motion of particles and light in the background described by Eq. (\ref{35}) 
can be studied by solving the geodesic equations,
\begin{equation}
\ddot{x}^\mu + \Gamma^\mu_{\alpha\beta} \dot{x}^\alpha\dot{x}^\beta = 0,
\label{36}
\end{equation}
where ``dot'' denotes derivative with respect to an arbitrary affine parameter $u$. In the case of 
particles $u$ is usually identified with the proper time $\tau$. 
Using background coordinates $(t,x,y,z)$ it is straightforward to show
that the geodesic equations yield
\begin{equation}
\frac{d^2\vec{r}}{dt^2} =  \frac{c^2\alpha}{z^3}\left(1-\frac{ \vec{v}\,{}^2 + v_z^2}{c^2}\right)\hat{z},
\label{a}
\end{equation}
where $\vec{r}=x\hat{x}+y\hat{y}+z\hat{z}$ and $\vec{v}\,{}^2=(d\vec{r}/dt)^2$.
As can be seen from Eq. (\ref{a}) there is no acceleration in the directions parallel to 
the plane boundary, so we can set $v_x=v_{0x}$ and $v_y=v_{0y}$. 
At this point it should be stressed that 
\begin{equation}
v_x^2+v_y^2+v_z^2\left(1-\frac{\alpha}{z^2}\right) \le c^2,
\label{cond}
\end{equation}
where the equality holds for light only.
Motion initially parallel and perpendicular to the boundary will be examined next.

\begin{figure}[!hbt]
\leavevmode
\centering
\includegraphics[scale = 0.8]{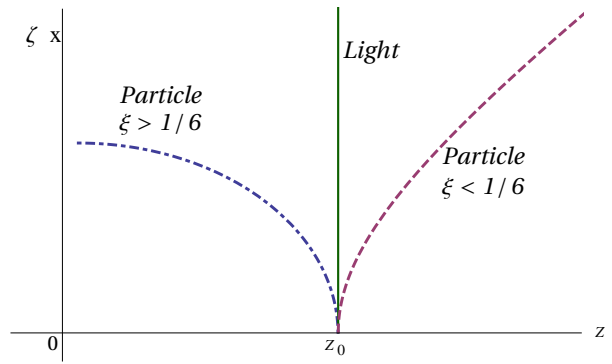}
\caption{{\small\sf (color online). Trajectories of particles and light with initial conditions 
$\vec{r}_0 = z_0 \hat{z}$ and $\vec{v}_0 = v_{0x}\hat{x}$. 
Light, in this special situation, is not affected by the nonuniform
energy-momentum density distribution. In order to make the quantum effects visible, 
$x$ has been rescaled by the factor $\zeta \doteq c|\alpha|^{1/2}/v_{0x}$.}}
\label{fig1}
\end{figure}

Implementing initial conditions $\vec{r} = z_0 \hat{z}$ and $\vec{v} = v_{0x} \hat{x}$ 
integration of Eq. (\ref{a}) shows
that a particle will travel with velocity ($v_{0x},0,v_z$), where
\begin{equation}
v_z^2 = c^2\alpha \left(1-\frac{v_{0x}^2}{c^2}\right)\left(\frac{1}{z_0^2} - \frac{1}{z^2}\right).
\label{58}
\end{equation}
Hence the particle will be subjected to an acceleration $a_z=  (1-v_{0x}^2/c^2)(c^2\alpha/z^3)$,
which can be positive or negative
depending on the sign of the parameter $\alpha$ defined by Eq. (\ref{20})
[note that according to Eq. (\ref{cond}) $v_{0x} < c$ for particles].
It will be accelerated in the $z$-direction, approaching the 
boundary or departing from it depending on whether
$\xi$ is greater or smaller than $1/6$, respectively.
These behaviors are depicted in Fig. \ref{fig1} where the dot-dashed 
curve corresponds to the trajectory of a particle in the case $\xi > 1/6$ 
and the dashed curve corresponds to the trajectory of a particle in the case $\xi < 1/6$. 
On the other hand, if a light signal is considered we must set $ds^2=0$ in Eq. (\ref{35})
and Eq. (\ref{a}) shows that the acceleration will be zero. Thus a light signal 
initially moving in the 
$xy$-plane will keep its initial velocity indefinitely: 
the trajectory of a light ray will be a straight line, as in Fig. \ref{fig1}, which  
is an interesting effect for the energy-momentum tensor in 
Eq. (\ref{5}) describes a nonuniform energy density distribution. 

Similar analysis can be done for the case of test particles initially moving in the $z$-direction.
In this case, considering initial conditions $\vec{r}=z_0\hat{z}$ and  $\vec{v}=v_{0z} \hat{z}$ in
Eq. (\ref{a}) implies that the velocity of particles is given by
\begin{equation}
v_z^2 = v_{0z}^2+ c^2\alpha \left(1-2\frac{v_{0z}^2}{c^2}\right)\left(\frac{1}{z_0^2} 
- \frac{1}{z^2}\right),
\label{58-b}
\end{equation}
with corresponding acceleration $a_z = (1-2v_{0z}^2/c^2)c^2\alpha/z^3$.
This acceleration will be positive 
or negative depending on the value of $\xi$ through the parameter $\alpha$
and also on the sign of the factor between brackets. 
We notice that if the initial velocity of the particle is such that $v_{0z}^2=c^2/2$
its acceleration will be zero and the particle will keep its initial velocity (up
to first order). 
However, if $v_{0z}^2$ is set to be greater (smaller) than $c^2/2$ the
particle will be affected by a negative (positive) acceleration in the case of $\alpha >0$ or the
opposite if $\alpha<0$.
If light is considered we obtain $v_z^2=c^2(1+\alpha/z^2)$ and 
$a_z= -c^2\alpha/z^3$ showing that its motion is not sensible to the
initial values $z_0$ and $v_{0z}$. 

%

%
\paragraph{ Gravitational spectral shift. ---}
Now let us concentrate only on the motion of light for which $ds^2=0$. 
Assume that a light signal is emitted near the boundary at ($0,0,z_A$)
and received at ($0,0,z_B$), with $z_B>z_A$. Standard calculations 
\cite{landau1951,foster2006} show that the gravitational potential $\phi$ due to the 
quantum field will induce a spectral shift in the light frequency $\nu$ given by
%
\begin{equation}
\frac{\Delta\nu}{\nu_A} = \frac{\phi_A-\phi_B}{c^2} = 
\frac{\alpha}{2}\left(\frac{1}{z_A^2}-\frac{1}{z_B^2}\right),
\label{90}
\end{equation}
where $\nu_i$ ($i=A,B$) is the frequency at $z_i$ and $\Delta\nu \doteq \nu_B-\nu_A$.
This spectral shift depends on the sign of the parameter $\alpha$. 
In the case of $\xi<1/6$ (which includes the minimal coupling)
$\Delta\nu/\nu_A > 0$, corresponding to a blue-shift, i.e., the light frequency 
received by an observer placed at $(0,0,z_B)$
is greater than the light frequency of the emitted signal. On the other hand, 
the case $\xi>1/6$ yields $\Delta\nu/\nu_A < 0$ which corresponds to a red-shift, i.e., the 
frequency of received light signal is smaller than the frequency of the emitted 
one. 
In principle, measurement of the spectral shift could set the 
value of the curvature coupling parameter.  
Clearly, no shift would appear for the conformal coupling.

At this point, only for pedagogical purposes, it is worthwhile investigating the numbers that would
come from a hypothetical experimental test if that were at all available. 
Using the known values for the physical constants that appear in Eq. (\ref{90}) we obtain
$\Delta\nu/\nu_A \approx 4 \times 10^{-72}(1-6\xi){\rm m}^2/z_A^2$. The magnitude of the
effect depends on the distance to the plane boundary that a light signal can be emitted
and also on the value of $\xi$. Shorter distances $z_A$ or stronger
curvature coupling produce larger spectral shift. 
Let us suppose we had at our disposal a massless scalar field at laboratory so that a 
measurement could be performed as described here. 
Just to illustrate the present accuracy to measure frequencies, the Laser 
Interferometer Gravitational-Wave Observatory (LIGO) is
able to measure spectral shift of the order of $10^{-21}$. Thus
without setting a specific value for the curvature coupling parameter (which could
be large), for a light signal initially at 
\begin{equation}
z_A \lesssim \sqrt{|1-6\xi|}
\times 10^{-25}{\rm m}, 
\label{lastbutone}
\end{equation}
departing from the plane boundary, the above effect would be detectable.

If a plane boundary with a non-negligible surface mass density $\sigma$ 
were considered, additional contributions to
the gravitational potential should be included in the above analysis.
Such contributions would come from classical and quantum levels. The 
dominant one comes from the classical gravitational potential 
$\phi_N(z) = 2\pi G\sigma z$, which leads 
to a Newtonian spectral shift given by 
\begin{equation}
\left(\frac{\Delta\nu}{\nu_A}\right)_{\mbox{\tiny\rm N}} = 
-\frac{2\pi G \sigma}{c^2}(z_B-z_A).
\label{last}
\end{equation}
This value may be compared with
the spectral shift caused by the Casimir stresses in the case of a
plane boundary with negligible $\sigma$, Eq. (\ref{90}). 
As one can see, smaller distances produce weaker
effect, in contrast with the quantum vacuum effect. 
Thus for small enough distances the Casimir-induced spectral shift dominates
over its Newtonian counterpart.

\paragraph{ Final remarks. ---}
Within the present-day laboratory limitations
and dealing with material boundaries, the effects described above would 
hardly be detectable except when $\xi$ is very large as Eq. (\ref{lastbutone}) suggests.
One of the main difficulties would be to find a substance to act as a
``perfect conductor'' for a massless scalar field $\varphi$ (assuming
that $\varphi$ is available).
Another difficulty in considering ordinary material surfaces is the
breaking of smoothness at atomic scales, where the idealization of
a plane material boundary does not apply. 
However, smoothness up to the Planck scale ($\approx 10^{-35}{\rm m}$) is expected to occur
in certain geometrical surfaces appearing in the context of cosmic defects. 
For instance, if a domain wall \cite{vilenkin1981} is to be considered the
Casimir spectral shift would be substantially enhanced. 

Before closing a word of caution is in order. The expression
for the expectation value of the energy-momentum tensor in Eq. (\ref{5})
should not be considered to hold arbitrarily close to the wall. 
When $\xi<1/6$ ($\xi>1/6$ if the Neumann boundary condition is taken),
the energy density in Eq. (\ref{5}) is negative, violating the weak energy condition, 
and suggesting that the effects described above might be of the 
same nature of those exotic phenomena that usually plague backgrounds
with unrestricted negative energies \cite{ford2010}.
This is not the case though. Boundary conditions on a 
geometrical plane are merely an artifact to simulate the presence
of a real wall. If instead of an idealized Dirichlet plane 
a real wall were taken into account (see, e.g., Ref. \cite{olum2003}), Eq. (\ref{5})
would  hold only in the bulk and the average null energy condition
would be satisfied (see also Ref. \cite{graham2005}). This fact is enough to prevent
some of the exotic phenomena mentioned before, namely, superluminal velocities
and appearance of time machines.

\begin{acknowledgments}
This work was partially supported by the 
Brazilian research agencies CNPq and FAPEMIG.
\end{acknowledgments}


\end{document}